\documentclass[aps,pre,showpacs]{revtex4}
\usepackage{graphicx}
\usepackage{wasysym}

\newcommand{\be}{\begin{equation}}
\newcommand{\ee}{\end{equation}}
\newcommand{\bea}{\begin{eqnarray}}
\newcommand{\eea}{\end{eqnarray}} 

\begin{document}

\title{Solution of voter model dynamics on annealed small-world networks}

\author{Daniele Vilone}
\email{daniele@pil.phys.uniroma1.it}
\affiliation{Dipartimento di Fisica, Universit\`a di Roma ``La Sapienza'',
and SMC-INFM, Unit\`a di Roma 1, P.le A. Moro 2, I-00185 Roma, Italy}
\author{Claudio Castellano}
\email{castella@pil.phys.uniroma1.it}
\affiliation{Dipartimento di Fisica, Universit\`a di Roma ``La Sapienza'',
and INFM, Unit\`a di Roma 1, P.le A. Moro 2, I-00185 Roma, Italy}

\date{\today}

\begin{abstract}
An analytical study of the behavior of the voter model on the small-world
topology is performed.
In order to solve the equations for the dynamics, we consider an annealed
version of the Watts-Strogatz (WS) network, where long-range connections
are randomly chosen at each time step.
The resulting dynamics is as rich as on the original WS network.
A temporal scale $\tau$ separates a quasi-stationary disordered
state with coexisting domains from a fully ordered frozen configuration.
$\tau$ is proportional to the number of nodes in the network, so that the
system remains asymptotically disordered in the thermodynamic limit.
\end{abstract}


\maketitle

\section{Introduction}
\label{Introduction}
The relationship between nontrivial interaction topologies
and ordering phenomena is still a largely unexplored topic.
The recent burst of activity on complex networks has revealed that many
technological, social and biological systems have interaction patterns
markedly different from structures traditionally studied as regular lattices
and random graphs~\cite{Barabasi02,Newman03}.
The interest is now focusing on how such complex
topologies affect dynamical processes taking place on them.
Models of ordering dynamics play an important role in this context, since
they are commonly used to study social phenomena, like cultural assimilation
and opinion
dynamics~\cite{Axelrod97,Castellano00,Sznajd-Weron00,Krapivsky03},
for which the interaction patterns are more plausibly described
by complex networks than by regular lattices.

Some ordering processes on complex networks have recently been considered,
including the zero temperature Glauber dynamics
of the Ising model on the Watts-Strogatz network~\cite{Boyer03} and
the Axelrod model on small-world and scale-free networks~\cite{Klemm03}.

In a recent paper~\cite{Castellano03}, we have studied numerically the dynamics
of the voter model on a small-world network.
This structure, more precisely the Watts-Strogatz (WS) network~\cite{Watts98},
is one of the simplest examples of complex topology. Depending on the parameter
$p$ (to be specified below) it interpolates between a one-dimensional
lattice with periodic boundary conditions (for $p=0$) and a
random graph (for $p=1$). It has been shown~\cite{Watts98}, that in a well
defined range of intermediate values of $p$, the network has, simultaneously,
global properties typical of random graphs (small average distance between
nodes) and local properties (clustering) typical of regular
structures.

The voter model is possibly the simplest model of an ordering
process~\cite{Liggett85}.
On each site a discrete variable $\sigma $ is defined, that may assume
two values ($\sigma =\pm 1$) representing two opposite options, for instance
the electoral choice in favor of two different candidates.
Starting from a disordered initial condition, the model follows a simple
dynamical evolution: at each time step one site is selected at random
and set equal to one of its nearest neighbors, chosen at random in its turn.
On regular lattices, in $d=1$ and $d=2$ the model converges to an ordered
state with all variables having the same value, whereas for $d \geq 3$
the system reaches a disordered stationary
state~\cite{Krapivsky92, Frachebourg96}.
The voter model on complete graphs has been considered
recently~\cite{Dickman02,Slanina03}.

In Ref.~\cite{Castellano03} we have found that the nontrivial connectivity
pattern of the WS network has a deep impact on the ordering dynamics of the
voter model.
In particular, after an initial transient, the system settles in a
quasi-stationary state with coexisting domains. If the system size is infinite
this state persists forever: at odds with naive expectations, long-range
connections prevent complete ordering from being reached.
If the system size is kept finite instead, the stationary state has a finite
lifetime and the fully ordered state is quickly reached at the end of it.
Interestingly, the dependence of the lifetime on the system size is such
that the ordered state is reached earlier than on a
one-dimensional lattice of the same size. This partially restores the
intuitive picture that long-range connections should speed up the ordering
process.

In this work we analyze the same problem from the analytical point
of view. The very simple form of the transition rates for the voter model
results in equations of motion for the correlation
functions that are not coupled with each other.
The only difficulty is to carry out the average over trajectories for
a fixed realization of the network and to average over the topology afterward.
To overcome this problem we use an annealed approximation, consisting
in averaging over the topology before averaging over the trajectories.
This approximation is exact for an effective network, which we call annealed
WS network, where long-range interactions are not quenched
from the beginning but are extracted randomly at each time step.
The dynamics of the voter model on the annealed WS network is then
solved, revealing a phenomenology as rich as in the quenched case.
By comparing the exact results on the annealed WS network with the numerical
simulations on the quenched one, it turns out that the only discrepancy
between the two cases is the dependence on the parameter $p$ of the
correlation length in the stationary disordered state.

The paper is organized as follows.
In the next section we define precisely the voter model dynamics and the
WS topology on which the process occurs.
Section~\ref{analytical} is devoted to the analytical solution of the
dynamics. The equation of motion for the correlation
function is derived, the annealed approximation is introduced and the
behavior of the system is studied.
In section~\ref{num} we check numerically the analytical results
for the annealed case and compare them with simulations of the
quenched case.
In section~\ref{extension} we extend the results to the case where
each site is initially connected to $2 \nu$ neighbors.
The final section contains a short discussion of the findings.

\section{The model}
\label{model}

We consider a small-world network defined as the superposition of a
one-dimensional lattice of $L$ sites with periodic boundary conditions
and a random graph~\cite{Monasson99}.
In general one can start from a lattice with each site linked to $\nu$
neighbors on the right and $\nu$ on the left. We now consider $\nu=1$,
deferring the discussion of generic values of $\nu$ to Section~\ref{extension}.
More precisely, site $i$ is initially connected with sites $i-1$ and $i+1$.
Then a link is added between any pair of non-nearest neighbor sites
with probability $p/L$. In this way the total number of
edges in the system is $L + L (L-3)/2 \cdot p/L$, so that the average
degree per site is finite ($2+p$) in the thermodynamic limit $L \to \infty$.
The generalization to an initial lattice with connections to $k$ nearest
neighbors is straightforward.
This topology slightly differs from the one originally introduced by Watts
and Strogatz~\cite{Watts98} but has the same properties and
is more amenable to analytical treatment.
The topology is fully specified by the adjacency matrix $Q(i,j)$, which
is 1 if $i$ and $j$ are connected and 0 otherwise.
The probability distribution of its elements is
\be
\label{pr}
P[Q(i,j)]=\left\{ \begin{array}{cc}
		\delta_{Q(i,j),0} & \mbox{for $i=j$} \\
		 & \\
		\delta_{Q(i,j),1} & \mbox{for $i=j\pm 1$} \\
		 & \\
		{p \over L} \delta_{Q(i,j),1}+(1-{p \over L})
                   \delta_{Q(i,j),0} 
                 & \mbox{otherwise.}
	\end{array}
	\right.
\ee

The voter model dynamics is defined by the transition rates.
If we call $\{ \sigma\} $ the spin configuration of the system, that is
$\{ \sigma\}=\{ \sigma_{1},\sigma_2 ,\dots ,\sigma_i ,\dots ,\sigma_L\}$
and we indicate with $\{ \sigma '\}_{i}$ the same configuration with the
$i$-th spin flipped, the transition rate from state $\{ \sigma\} $ to state
$\{ \sigma '\}_{i}$ is, in complete analogy with the definition for regular
lattices,
\be
w(\{ \sigma \} \rightarrow \{ \sigma '\}_{i})= \frac{1}{2}
\left(1-\frac{\sigma_i}{z_i} \sum_k Q(i,k) \sigma_k \right),
\label{ratesw}
\ee
where $z_i$, the degree of site $i$, is
\[
z_j = \sum_i Q(i,j).
\]

\section{Analytical treatment}
\label{analytical}

\subsection{Equation for the correlation function}

Given the explicit expression~(\ref{ratesw}) of the transition rates,
from the master equation
\be
{d \over dt} P(\{\sigma \},t) =
\sum_i w(\{ \sigma' \}_{i} \rightarrow \{ \sigma\}) P(\{\sigma' \}_i,t) -
\sum_i w(\{ \sigma \} \rightarrow \{ \sigma '\}_{i}) P(\{\sigma \},t),
\ee
one can derive~\cite{Glauber63, Rednerbook} the equation of motion
for the mean spin at site
$j$, $s_j \equiv \langle \sigma_j \rangle$
\be
{d s_j \over dt} = - s_j + \sum_k {Q(j,k) s_k \over z_j},
\label{meanspin}
\ee
and for the two-point correlation function $C_{j,k} \equiv
\langle \sigma_j \sigma_k \rangle$
\be
\frac{d C_{j,k}}{d t}= -2 C_{j,k}+
\sum_i {Q(j,i) C_{k,i} \over z_j}+
\sum_i {Q(k,i) C_{j,i} \over z_k}.
\label{corr}
\ee
For $p=0$, Eqs.~(\ref{meanspin}) and~(\ref{corr})
coincide with the equations for the one-dimensional voter
model~\cite{Frachebourg96}.

In order to study the dynamics of the voter model on the Watts-Strogatz
network we must average Eqs.~(\ref{meanspin}) and~(\ref{corr}) over
the disordered topology. Indicating with an overbar averaged quantities
$\overline{A}=\int \prod_{i,j} dQ(i,j) A \cdot P[Q(i,j)]$, the
equation for the mean spin is
\be
{d \overline{s_j} \over dt} = - \overline{s_j} + \sum_k
\overline {\left( {Q(j,k) s_k \over z_j} \right)},
\label{meanspinav}
\ee
and the one for the pair correlation function is
\be
\frac{d \overline{C_{j,k}}}{d t}= -2 \overline{C_{j,k}}+
\sum_i \overline{\left( {Q(j,i) C_{k,i} \over z_j} \right)} +
\sum_i \overline{\left( {Q(k,i) C_{j,i} \over z_k} \right)}.
\label{corrav}
\ee
To evaluate the average values appearing on the r. h. s. of
Eqs.~(\ref{meanspinav}) and~(\ref{corrav}) we introduce the annealed
approximation
\be
\overline{\left(\frac{Q(j,k)}{z_j}\right)\cdot A}=
\overline{\left(\frac{Q(j,k)}{z_j}\right)}\cdot \overline{A}.
\label{annealed}
\ee
This approximation can be seen as considering annealed transition rates
\be
\overline{w}(\{ \sigma \} \rightarrow \{ \sigma '\}_{i})= \frac{1}{2}
\left[1-\sum_k \overline{\left({Q(i,k) \over z_i}\right)}
\sigma_i \sigma_k \right].
\label{rateswannealed}
\ee
The evaluation of $\overline{\left(\frac{Q(j,k)}{z_j}\right)}$ is readily
performed
\begin{eqnarray}  \label{ar}
\overline{\left(\frac{Q(j,k)}{z_j}\right)} & = &
\int \prod_{l,m} dQ(l,m) P[Q(l,m)] \frac{Q(j,k)}{\sum_i Q(j,i)} = \nonumber \\
				 	   & = &
\int dQ(j,k) P[Q(j,k)] Q(j,k) \int \prod_{l,m\neq j,k} dQ(l,m)
P[Q(l,m)] \frac{1}{\sum_i Q(j,i)}.
\end{eqnarray}

Inserting Eq.~(\ref{pr}) in Eq.~(\ref{ar}), one easily finds for $k = j$
\be
\overline{\left(\frac{Q(j,j)}{z_j}\right)}=0,
\ee
for $k = j \pm 1$
\be
\overline{\left(\frac{Q(j,j\pm 1)}{z_j}\right)}=
\int dQ(j,k) P[Q(j,k)] Q(j,k) \sum_{n=0}^{L-3}
\left(\begin{array}{c} L-3 \\ n \end{array}\right)
\frac{(p/L)^n (1-p/L)^{L-3-n}}{1 + Q(j,k) + n} = f_2(p/L,L-3),
\ee
and similarly, for $k\neq j,j\pm1$, 
\be
\overline{\left(\frac{Q(j,k)}{z_j}\right)}= {p \over L} \cdot f_3(p/L,L-4),
\ee
where
\be
f_R (\alpha,N)=\sum_{n=0}^{N}\left(\begin{array}{c} N \\ n \end{array}\right)
\frac{\alpha^n (1-\alpha)^{N-n}}{R+n}.
\ee
Explicit formulas for the functions $f_2$ and $f_3$ are given in the Appendix.
In the limit of large $L$ they tend to the simple forms
$$
F_2(p) = \lim_{L \to \infty} f_2(p/L,L-3) = {1 \over p} - 
{(1-e^{-p}) \over p^2}
$$
and
$$
F_3(p) = \lim_{L \to \infty} f_3(p/L,L-4) = {1 \over p} - {2 \over p^2} +
{2(1-e^{-p}) \over p^3},
$$
that go to $1/2$ and $1/3$, respectively, in the limit $p \to 0$.

The dynamical rule corresponding exactly to rates~(\ref{rateswannealed})
is easily found.
At each time step one site (site $i$) of a one-dimensional lattice is
randomly selected.
Then, with probability $f_2(p/L,L-3)$ one of the two nearest neighbors
($i+1$ or $i-1$) is chosen and $\sigma_i$ is set equal to it.
With probability $(p/L) f_3(p/L,L-4)$ instead, one randomly chooses one of
the other $L-3$ sites and sets $\sigma_i$ equal to it.
In this way, the effective topology over which the dynamics takes place changes
at each time step. We call such network ``annealed'' WS network.
The average properties are those of the original ``quenched'' WS network,
but no permanent connection exists between non nearest neighbor sites.

We can now write down the explicit form of the equations of motion for the
mean spin and the correlation function on the annealed Watts-Strogatz network
\be
\dot{s_j}=-s_j+f_2 (p/L,L-3)\ (s_{j+1}+s_{j-1})+{p \over L} \
f_3 (p/L,L-4)\sum_{l\neq j,j\pm 1}s_l
\label{meq1}
\ee
\begin{eqnarray}  \label{meq2}
\dot{C}_{j,k} & = & -2C_{j,k}+f_2 (p/L,L-3)\ 
(C_{j+1,k}+C_{j-1,k}+C_{j,k+1}+C_{j,k-1}) + \nonumber \\
	      & + & {p \over L} \ f_3 (p/L,L-4) 
\left(\sum_{l\neq j,j\pm 1} C_{k,l} + \sum_{l\neq k,k\pm 1} C_{j,l}\right).
\end{eqnarray}
For simplicity, here and in the following, the overbar is omitted.
Also the arguments of $f_2$ and $f_3$ will often be omitted.

Eq.~(\ref{meq2}) is complemented by the boundary condition
$C_{j,j}=1$ and by the initial condition $C_{j,k}(t=0)$.
We consider an initial fully uncorrelated state
$C_{j,k}(t=0)=\delta_{j,k}$.
Hence the correlation function depends only on $r=|j-k|$ for all times and
the equation of motion is
\be
\dot{C}(r)=-2 \left(1 + {p f_3 \over L} \right) C(r) + 
2 \left( f_2 - {p f_3 \over L} \right)
\left[C(r+1) + C(r-1)\right] + 2 {p f_3 \over L} \sum_{l=0}^{L-1} C(l).
\label{meq3}
\ee
where the relation $2 f_2 (p/L,L-3) +(L-3) p/L f_3 (p/L,L-4)-1=0$, proved
in the Appendix [Eq.~(\ref{eqapp})], has been used.

If we sum Eq.~(\ref{meq1}) over $j$ and divide by $L$,
we obtain the temporal evolution of average total magnetization
$M= (1/L) \sum_j s_j$:
\be
{d M \over dt} = \left[-1+\ 2 f_2 (p/L,L-3) +(L-3)\ p/L \ f_3(p/L,L-4)\right]M.
\label{M1}
\ee
Since the coefficient on the right hand side vanishes [Eq.~(\ref{eqapp})]
the average total magnetization is conserved, as in the dynamics on regular
lattices.

\subsection{The stationary state}
\label{stationary}

We now turn to the analysis of Eq.~(\ref{meq3}) and consider first the
continuum limit in real space, yielding
\be
\dot{C}=-2 \left(1-2 f_2 \right) \left(1 + {3 \over L-3} \right) C + 
2 \left( f_2 - {1- 2 f_2 \over L-3} \right) C'' +
2 {1 - 2 f_2 \over L-3} \int_0^{L} dr C(r).
\label{eqcont}
\ee

Let us look for final configurations of the system, i. e. solutions
of the stationary equation
\be
C''-\lambda_L^2 C +\Phi_L = 0,
\label{rsec}
\ee
where
\be
\lambda_L^2 = {(1-2 f_2) [1 + 3/(L-3)] \over [f_2 - (1 -2 f_2)/(L-3)]} > 0
\ee
and
\be
\Phi_L = {\lambda_L^2 \over L} \int_0^{L} dl C(l).
\label{Phy}
\ee
For finite $L$, taking into account that $C(r=0)=1$ and
$|C(r)|\leq 1\ \forall r$, one obtains
\be
C(r)=\frac{\Phi_L}{\lambda_L^2}+
\left[1-\frac{\Phi_L}{\lambda_L^2}\right]
\exp [-\lambda_L r].
\label{primasol}
\ee
Imposing the consistency of Eq. (\ref{primasol}) with Eq. (\ref{Phy}),
the final correlation function turns out to be
\be
C(r)=1\ \ \ \ \ \ \ \ \ \  \forall r \geq 0.
\label{firstres}
\ee
The final configuration of the system is a frozen fully ordered state.

On the other hand, in the thermodynamic limit $L \to \infty$,
$1/L \int_0^L dr C(r)=C(\infty)$ so that the solution is
\be
C(r) = C(\infty) + \left[1- C(\infty) \right] \exp [-\lambda(p) r],
\ee
with $\lambda^2(p) = \lim_{L \to \infty} \lambda_L^2$.
From Eq.~(\ref{eqcont}) one obtains $\dot{C}(\infty) = 0$ and since
$C(\infty)=0$ for $t=0$ we have
\be
C(r)=\exp[-\lambda(p) r].
\label{secondres}
\ee
Hence the stationary configuration of the system is disordered,
with a correlation length $\xi_p$
\be
\xi_p=\frac{1}{\lambda(p)} = {1 \over \sqrt{1/F_2(p)-2}} = 
{1 \over \sqrt{p^2/(p-1 + e^{-p})-2}},
\label{csi}
\ee
where the explicit form of $F_2(p) = \lim_{L \to \infty} f_2(p/L,L-3)$
is computed in the Appendix.
In the limit of small $p$ the correlation length diverges as $p^{-1/2}$.

\subsection{Preasymptotic dynamics}

We now study the equation of motion~(\ref{meq3}) in Fourier space
by closely following the treatment for the Ising model
on a one-dimensional lattice~\cite{Bray89}.
In this way, not only the stationary state but also the
preasymptotic dynamics can be analyzed.
Introducing
\be
C(r,t)=\frac{1}{L}\sum_{k'} c_{k'}(t) e^{ik'r},
\ee
with $k' = 2 \pi n/L$, $n=-L/2,...,L/2$,
multiplying both sides of Eq.~(\ref{meq3}) by $e^{-ikr}$,
summing over $r$ from $1$ to $L-1$, we obtain
$$
\frac{1}{L}\sum_{k'} \dot{c}_{k'}(t) \sum_{r=1}^{L-1} e^{i(k'-k)r} =
- 2 A \frac{1}{L}\sum_{k'} c_{k'}(t) \sum_{r=1}^{L-1} e^{i(k'-k)r}
+ 2 B \frac{1}{L}\sum_{k'} c_{k'}(t) (e^{i k'} + e^{-i k'})
\sum_{r=1}^{L-1} e^{i(k'-k)r}
+ 2 {p \over L} f_3 C_0 \sum_{r=1}^{L-1} e^{-i k r}
$$
Using
$\sum_{r=1}^{L-1} e^{i(k-k')} = L \delta_{k,k'}-1$, we get
\be
\dot{c}_k(t)=-\gamma_k c_k(t)+A(t)+2 {p f_3 \over L} c_0(t) (L\delta_{k,0}-1),
\label{trans2}
\ee
where 
\be
\gamma_k= 2 \left[1+ {p f_3 \over L}-2 \left(f_2-  {p f_3 \over L}\right)
\cos(k) \right],
\ee
and
\be
A(t)=\frac{1}{L}\sum_k c_k(t) \gamma_k.
\ee
The disordered initial condition implies $c_k(t=0)=1$ for all $k$.
The boundary condition $C(r=0,t)=1$ implies
\be
1 = {1 \over L} \sum_k c_k(t)
\label{bc}
\ee
for all $t$.
Eq.~(\ref{trans2}) can be solved by Laplace transform methods. Introducing
\be
\hat{c}_k(s)=\int_0^{\infty}c_k(t)e^{-st} dt.
\ee
Eq.~(\ref{trans2}) becomes
\be
s \hat{c}_k(s) - 1 = - \gamma_k \hat{c}_k(s) + \hat{a}(s) +
2 p f_3 \hat{c}_0(s) (\delta_{k,0}- 1/L),
\label{eqhata}
\ee
where $\hat{a}(s)=L^{-1}\sum_k \gamma_k \hat{c}_k(s)$ is the Laplace transform
of $A(t)$.
The coefficients $\hat{c}_k(s)$ can be formally written down
\be
\hat{c}_k(s)=\left\{ \begin{array}{cc}
	\frac{\hat{a}(s)+1}{s+\gamma_0/L} & \ \ k=0 \\
		& \\
	\frac{\hat{a}(s)+1-\gamma_0 \hat{c}_0(s)/L}{s+\gamma_k} 
        & \ \ k\neq 0. \\
	\end{array}
	\right.
\label{trans4}
\ee
Notice that $\gamma_0 = 2 p f_3$.
To determine $\hat{a}(s)$ we transform the boundary
condition~(\ref{bc}), use the expressions~(\ref{trans4})
and replace the discrete sum over $k$ with an integral (which is correct
in the limit of large $L$ we are interested in), obtaining
\be
{1 \over s} = {\hat{a}(s)+1 \over Ls+\gamma_0} +
\left(1 - {\gamma_0 \over Ls + \gamma_0} \right) [\hat{a}(s)+1]
\int_{-\pi}^{\pi} {dk \over {2 \pi}} {1 \over s+\gamma_k}.
\label{eqint}
\ee
For large $L$ we can take $\gamma_k(L) \to \gamma_k(L=\infty) =
2[1-(1-\gamma_0/2) \cos(k)]$.
The integral is easily carried out yielding
$[s^2+4s+4[1-(1-\gamma_0/2)^2]]^{-1/2}$.
We are interested in
$t \gg 1/p \Rightarrow s \ll \gamma_0$ so that we obtain
\be
{1 \over s} = [\hat{a}(s)+1] \left[ {1 \over Ls+\gamma_0} +
{1 \over 2 \sqrt{\gamma_0}}
\left(1 - {\gamma_0 \over Ls + \gamma_0} \right) \right].
\label{eqhata2}
\ee
From Eq.~(\ref{eqhata2}), one realizes the existence of a temporal scale
$\tau=L/\sqrt{\gamma_0}$ separating two different regimes.

For $t \ll \tau$, $\hat{a}(s) + 1 = 2 \sqrt{\gamma_0}/s$ so that
$c_0(t)=2 \sqrt{\gamma_0} t$ and $c_k(t)=2 \sqrt{\gamma_0}/\gamma_k$.
The correlation function is
\be
C(r,t) = 2 \sqrt{\gamma_0} {t \over L} + 
2 \sqrt{\gamma_0} \int_{-\pi}^{\pi} {dk \over 2 \pi}
{e^{i k r} \over 2 [1-(1-\gamma_0/2) \cos(k)]}.
\ee
The first term is negligible because $t \ll \tau$. Hence $C(r)$ does not
depend on time and decays exponentially with $r$.
This is a quasi-stationary state with coexisting domains
of fixed size. Such a regime lasts for a time $\tau$ which diverges for
$L \to \infty$, so that in the thermodynamic limit the system remains
asymptotically in this disordered state.

For $t \gg \tau$ instead, all $\hat{c}_k(s)$ vanish
except $\hat{c}_0(s) = L/s$.
Hence $C(r)=1$ and the system becomes completely ordered.
This is the asymptotic regime for a system of finite size.

Let us summarize the results obtained in this section.
We have solved the dynamics of the voter model on an annealed small-world
topology: in this way we have found that finite systems remain
in a disordered state for a time proportional to their size, and then
converge to the totally ordered configuration;
infinite systems instead reach a disordered final stationary state,
with a correlation function that decays exponentially over a distance $\xi_p$.
For small values of $p$ the correlation length $\xi_p$ diverges as $p^{-1/2}$.

\section{Numerical results}
\label{num}

In order to validate the analytical results presented above we have
performed numerical simulations of the voter model both on the annealed
and on the quenched Watts-Strogatz topology.

\begin{figure}
\includegraphics[angle=0,width=9cm,clip]{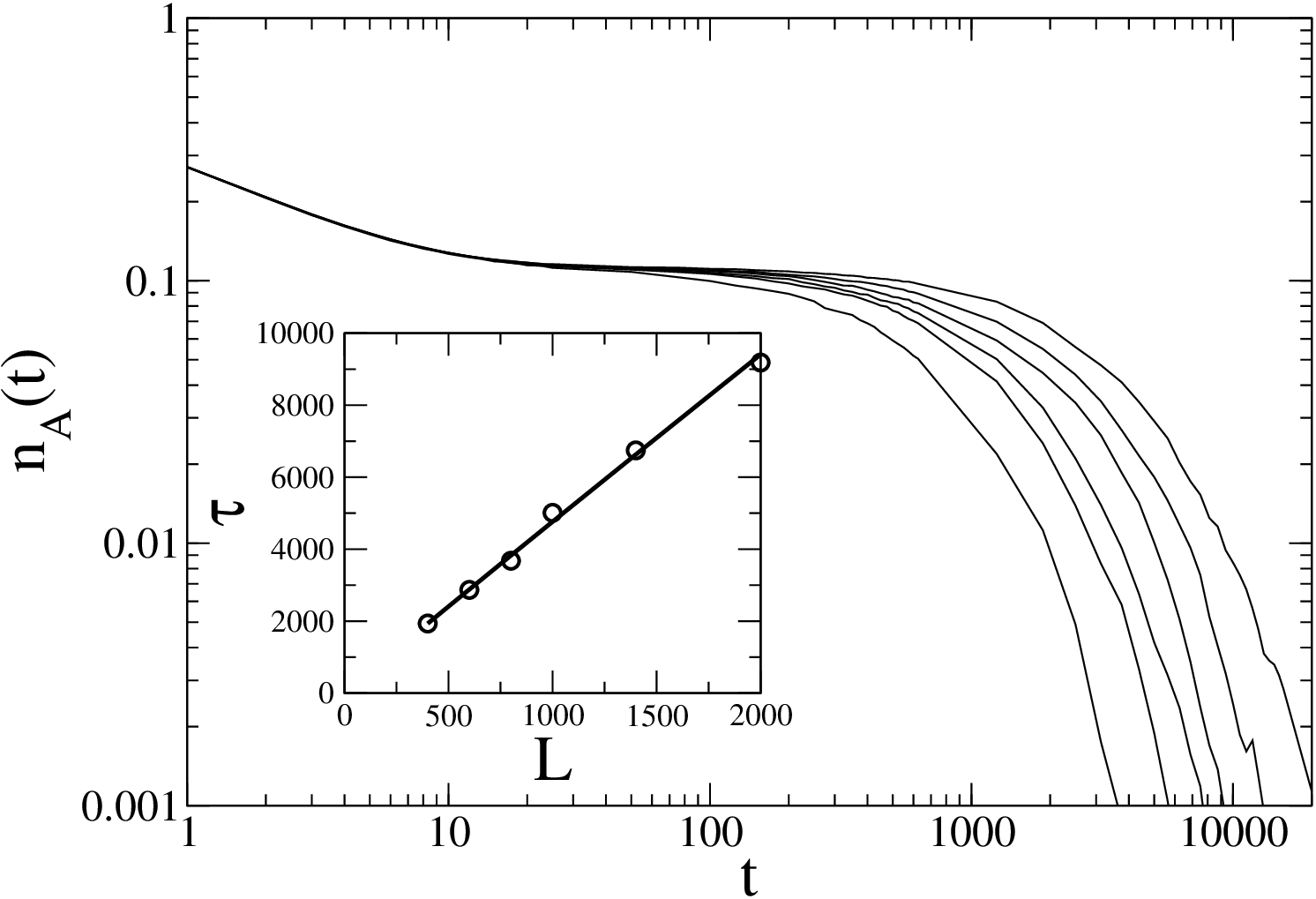}
\caption{Main: Plot of the fraction of active bonds $n_A$ for $p=0.1$ and
system sizes $L$ 400, 600, 800, 1000, 1400, 2000 (from left to right)
for the annealed case. Data are averaged over 1000 different realizations.
Inset: The duration $\tau$ of the plateau
in the main part of the figure (symbols) plotted versus $L$.
$\tau$ is evaluated as the time at which $n_A$ drops below 0.01.
The straight line is a power-law regression with best-fit exponent
equal to $0.98 \pm 0.03$.}
\label{Fig1}
\end{figure}

Figure~\ref{Fig1} reports, for the annealed case, the temporal behavior
of the fraction $n_A$ of active bonds, i. e. the fraction of nearest
neighbor sites with opposite values
of $\sigma$, for $p=0.1$ and several values of $L$.
After an initial decrease, typical of one-dimensional systems, a plateau
sets in. The analytical treatment predicts such a preasymptotic regime
to last for an interval proportional to $L$. The inset of Fig.~\ref{Fig1}
confirms the analytical finding.

\begin{figure}
\includegraphics[angle=0,width=9cm,clip]{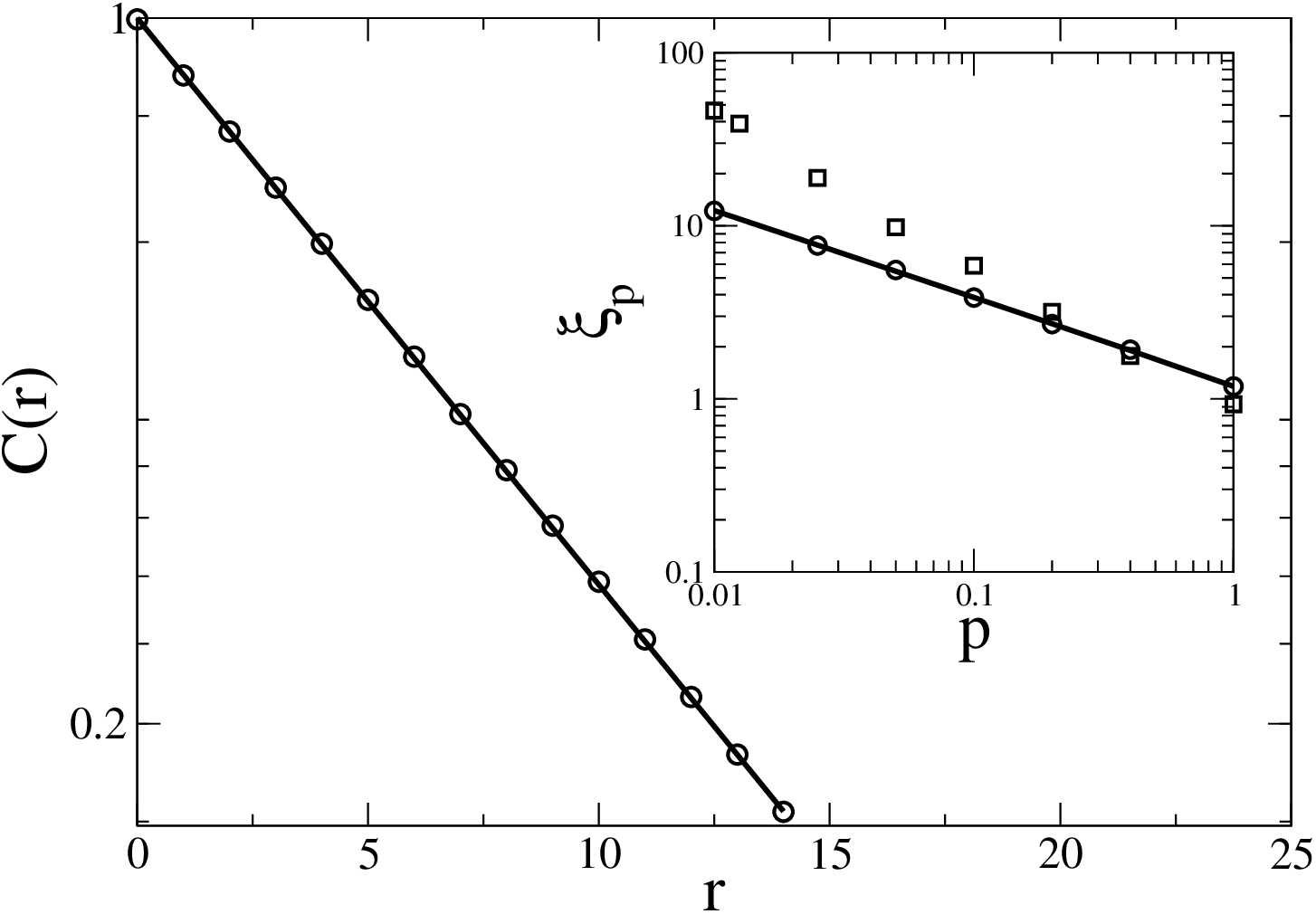}
\caption{Main: Correlation function $C(r)$ in the quasi-stationary state
for a system of size $L=10^5$ and $p=0.025$ (symbols), compared with
the analytical prediction, Eq.~(\ref{secondres}) (solid line).
Only one realization of the noise is considered.
The agreement is excellent.
Inset: The correlation length $\xi_p$ obtained numerically
from the decay of the correlation function $C(r)$ in the quasi-stationary
state of annealed WS networks with $L=10^5$ (circles),
compared with the analytical prediction, Eq.~(\ref{csi}) (solid line),
and the same quantity computed for the quenched WS network (squares).}
\label{Fig2}
\end{figure}

In the limit $L \to \infty$, the disordered state corresponding to the
plateau becomes the asymptotic one. The analytical solution predicts
an exponential form of the correlation function [Eq.~(\ref{secondres})]
with the correlation length $\xi_p$ given by Eq.~(\ref{csi}).
This analytical form is a very good approximation for the correlation
function also for very large, but finite, values of $L$ in the
quasi-stationary regime ($t<\tau$).
This is shown numerically in the main part of Fig.~\ref{Fig2}.
In the inset of the same figure we report the values of $\xi_p$
obtained numerically, which perfectly coincide with Eq.~(\ref{csi}).

\begin{figure}
\includegraphics[angle=0,width=9cm,clip]{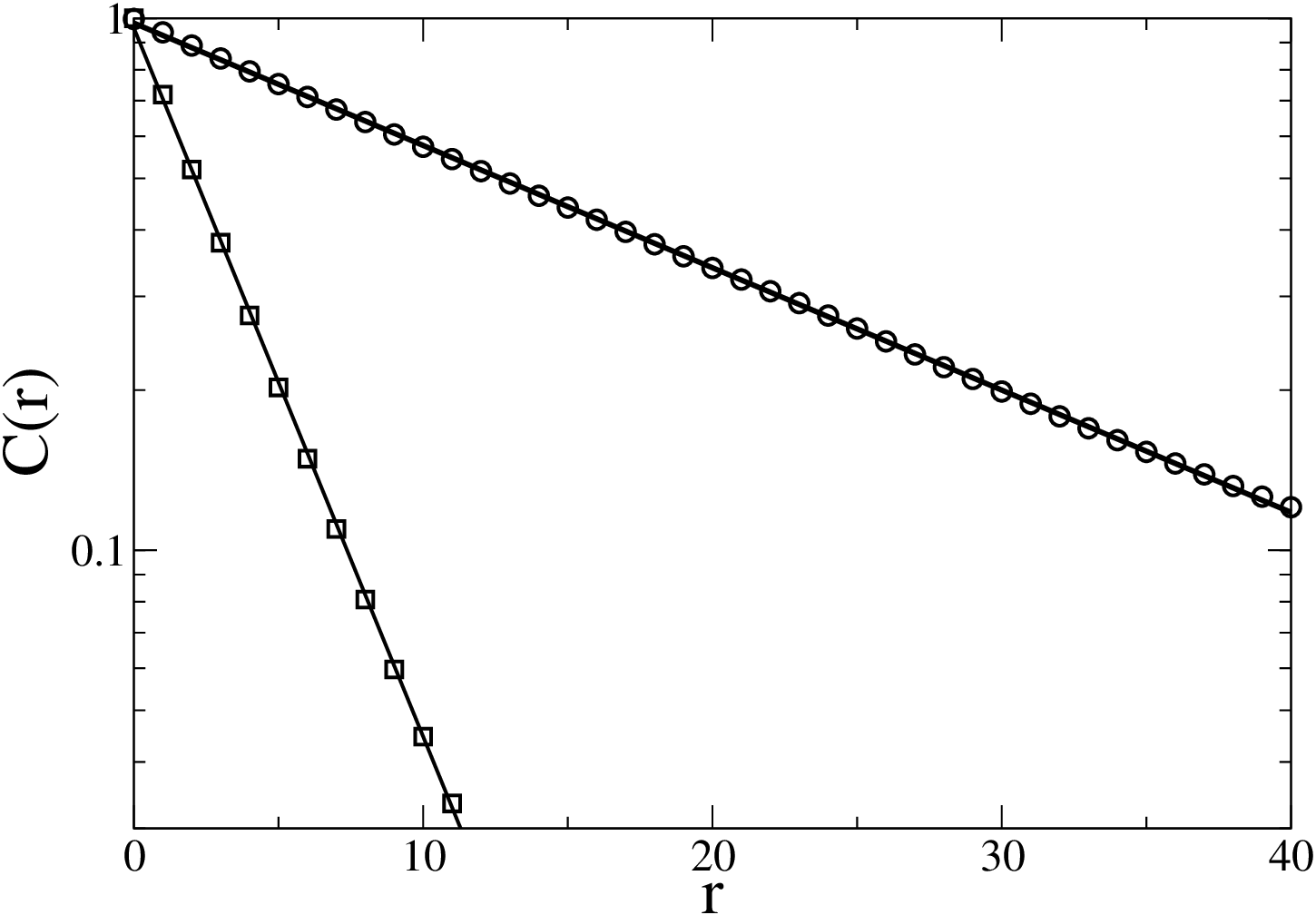}
\caption{
Correlation function $C(r)$ in the quasi-stationary state
for a quenched WS network of size $L=10^5$, $p=0.025$ (circles)
and $p=0.2$ (squares). Only one realization is considered.
The solid lines are exponential fits.}
\label{Fig3}
\end{figure}

We have introduced the annealed version of the Watts-Strogatz network
as an approximation for the WS network with quenched topology.
Hence it is interesting to compare the results of the two cases to
understand how well the approximation captures the behavior of the original
system.
In Ref.~\cite{Castellano03} we have already performed a numerical
investigation of the voter dynamics on the quenched WS topology.
By comparing the results reported in Ref.~\cite{Castellano03} and the
theoretical approach presented here, we see that the annealed approximation
correctly reproduces many of the important features of the original system,
i. e. the existence of a regime with a disordered state before
full order sets in and the linear dependence on $L$ of the temporal scale
$\tau$ between them.
Concerning the shape of the final correlation function in the case of infinite
quenched networks, Fig.~\ref{Fig3} shows that $C(r)$ is
exponential, another feature that is the same in the annealed
and the quenched cases.
What the annealed approximation is not able to capture is the
quantitative dependence of the correlation length on $p$. This is
shown in the inset of Fig.~\ref{Fig2}, where it is clear that, for small $p$,
$\xi_p\sim p^{-1}$ in the quenched system~\cite{Castellano03}, while
$\xi_p\sim p^{-1/2}$ in the annealed case.

\section{Extension to $\nu>1$}
\label{extension}
\begin{figure}
\includegraphics[angle=0,width=9cm,clip]{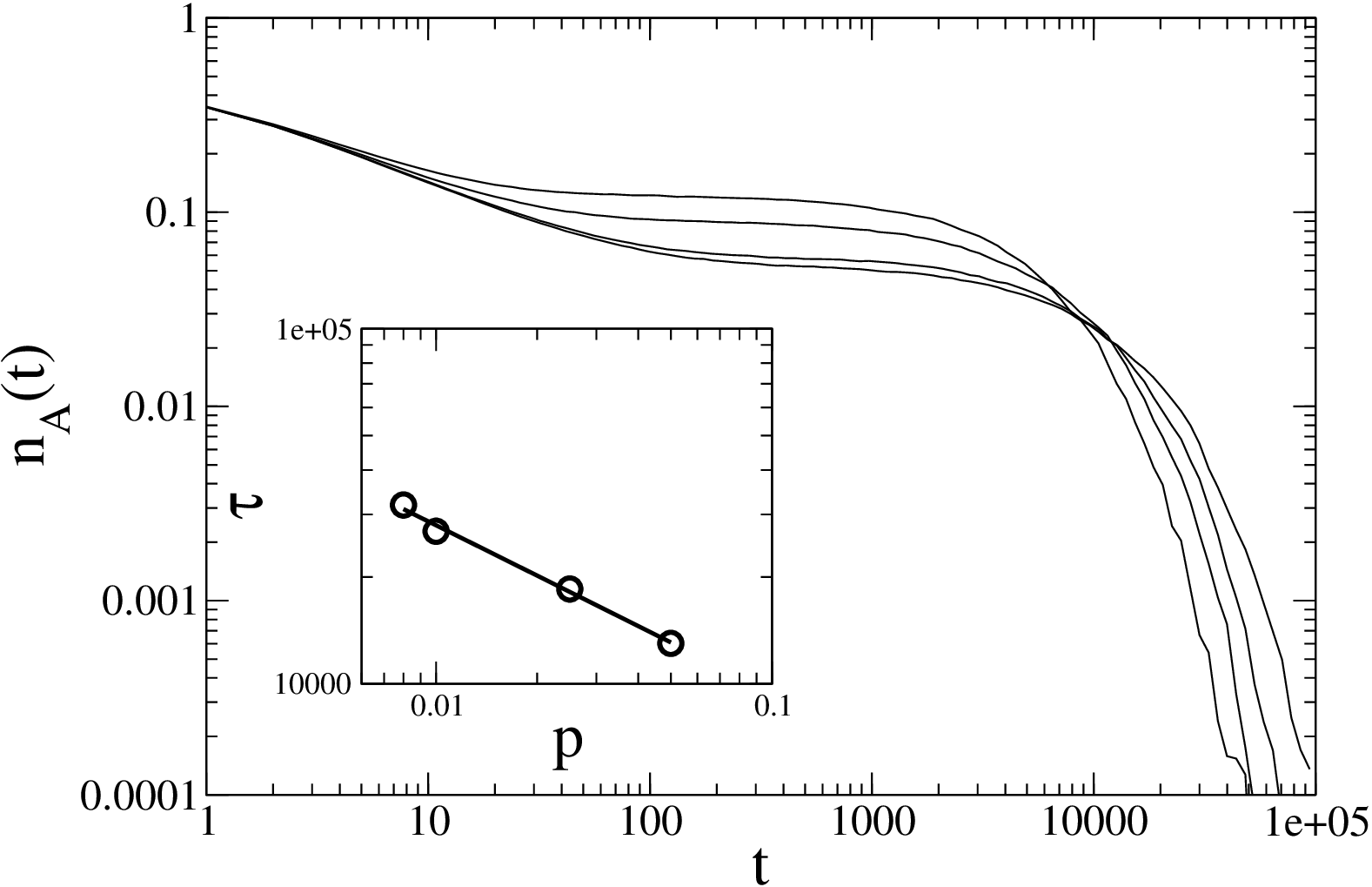}
\caption{Main: Plot of the fraction of active bonds $n_A$ for $L=3000$ and
$p=0.05$, $p=0.025$, $p=0.01$ and $p=0.008$ (from top to bottom)
for the annealed case. Data are averaged over 1000 different realizations.
Inset: The duration $\tau$ of the plateau
in the main part of the figure (symbols) plotted versus $p$.
$\tau$ is evaluated as the time at which $n_A$ drops below one tenth
of the value during the plateau.
The straight line is a power-law regression with best-fit exponent
equal to $0.47 \pm 0.03$.}
\label{Fig4}
\end{figure}

In the previous sections we have considered an initial one-dimensional
lattice with each site connected only to its nearest neighbors.
When the number of neighbors connected to each site is $\nu>1$,
one can quite easily extend the analytical calculations presented above.
On physical grounds one expects the qualitative picture to remain the same.
Also formulas are very similar. In general they involve, in the place of
$f_2$ and $f_3$ the functions $f_{2 \nu} (p/L,L-1-2 \nu)$ and
$f_{2 \nu+1} (p/L,L-2-2 \nu)$.
The equation for the average total magnetization becomes
\be
{d M \over dt} = \left[-1+\ 2 \nu f_{2 \nu} (p/L,L-1-2 \nu) 
+(L-1-2\nu)\ p/L \ f_{2 \nu+1}(p/L,L-2-2\nu)\right] M,
\label{M2}
\ee
and $M$ is again conserved due to the vanishing of the coefficient on the
right hand side [Eq.~(\ref{eqapp})].
The equation for the correlation function $C(r)$ becomes
\be
\dot{C}(r)=-2 \left(1 + {p f_{2 \nu+1}\over L}\right) C(r) + 
2 \left( f_{2 \nu} - {p f_{2\nu+1} \over L} \right)
\sum_{i=1}^{\nu} \left[C(r+i) + C(r-i)\right] +
2 {p f_{2 \nu+1} \over L} \sum_{l=0}^{L-1} C(l).
\label{Cnu}
\ee
Considering the continuum limit of Eq.~(\ref{Cnu}) and looking for the
stationary state one obtains
\be
C''-\lambda_L^2 C +\Phi_L = 0,
\ee
where now
\be
\lambda_L^2 = {(1-2 \nu f_{2 \nu}) [1 + (2 \nu+1)/(L-2 \nu-1)] 
\over [f_{2 \nu} - (1 -2 \nu f_{2 \nu})/(L-2 \nu-1)] \omega_{\nu}} > 0,
\ee
and
\be
\omega_{\nu} = \sum_{j=1}^{\nu} j^2 = {\nu (\nu+1) (2 \nu+1) \over 6}.
\ee
Again the only solution for finite $L$ is the completely ordered state
$C(r)=1$, while in the thermodynamic limit $C(r)=\exp[-\lambda(p) r]$.
The correlation length is
\be
\xi_p = 1/\lambda(p) =  {1 \over \sqrt{1/F_{2 \nu}(p)-2 \nu}},
\ee
where $F_{2 \nu}(p) = \lim_{L \to \infty} f_{2 \nu}(p/L,L-2\nu-1)$.
In the limit of small $p$, by expanding Eq.~(\ref{inte}), one finds
\be
\xi_p \sim (2 \nu p)^{-1/2}.
\ee
For what concerns the preasymptotic dynamics, the only formal change
in Eq.~(\ref{trans2}) is that $f_3$ is replaced by $f_{2 \nu+1}$, but
now the form of $\gamma_k$ is different
\be
\gamma_k= 2 \left[1+ {p f_{2 \nu+1} \over L}-2 \left(f_{2 \nu} - 
{p f_{2 \nu+1} \over L}\right) \sum_{j=1}^{\nu} \cos(j k) \right].
\ee
By applying the Laplace transform one gets an equation formally
equal to Eq.~(\ref{eqint}). For generic $\nu$ we cannot perform explicitly
the integral appearing in Eq.~(\ref{eqint}) and hence we cannot write
down the analogue of Eq.~(\ref{eqhata2}). However, we can guess that
the only change will be the replacement of the factor $1 /2\sqrt{\gamma_0}$
with some other factor independent from $L$.
Therefore the existence of two regimes separated by a temporal scale $\tau$
proportional to $L$ will be preserved.
Moreover, on physical grounds, we expect 
the proportionality factor to scale as $p^{-1/2}$ also for generic $\nu$.
To confirm this, we have performed numerical simulations of a system
with next-nearest neighbor connections ($\nu=2$).
The results, presented in Fig.~\ref{Fig4}, confirm the expectation.
The temporal scale $\tau$ separating the quasi-stationary disordered state
from the totally ordered configuration scales as $L p^{-1/2}$.
We can conclude that the behavior of the voter model on the small world
topology is qualitatively the same, regardless of the number $\nu$
of connections between neighbors.

\section{Conclusions}

In this paper we have studied analytically the voter model on the small-world 
topology.
We have considered an annealed version of the Watts-Strogatz network, where
long-range connections are not fixed, but chosen randomly at each time step.
In this way each realization of the voter model dynamics takes place in an
effective average small-world topology and this allows the exact solution
of the equation for the correlation function of the system.

The dynamical behavior of the model on the annealed topology is very
similar to the behavior on the quenched WS network.
Systems of finite size converge asymptotically to a totally ordered
frozen state after an intermediate quasi-stationary stage, characterized
by a finite correlation length. The duration of this 
preasymptotic regime is proportional to the number of sites.
Hence systems of infinite size never reach the ordered state and remain
in a disordered stationary state with finite domain size.
All these features are exactly the same both on the annealed and on the
quenched versions of the network.
A quantitative difference arises only in the dependence of the correlation
length $\xi_p$ on the probability $p$ of having a long-range connection.
This discrepancy is not surprising, since
a similar disagreement between the annealed and the quenched case has been
noted previously for diffusion on Watts-Strogatz networks~\cite{Lahtinen02}
and the voter model is well known to be related with first passage properties
of random walkers~\cite{Rednerbook}.
In that case, the crossover time separating short and long time behavior
of the mean number of distinct sites visited scales as $p^{-2}$ in the
quenched network and as $1/p$ in the annealed one.
This difference is due to the fact that in the quenched case a walker has to
diffuse over a distance $1/p$ before reaching a shortcut and deviating
from the one-dimensional behavior. This clearly requires a time $p^{-2}$.
In the annealed case instead, the time
needed to perform a long-range jump scales as $1/p$.
In the voter model on small-world topology the boundaries between ordered
domains perform a one-dimensional random walk for short times. The
behavior changes when the walkers make a long range jump.
As mentioned above, this requires a time $1/p$ in the annealed network
and $p^{-2}$ in the quenched one. This difference generates the different
scaling of the correlation length in the two types of topology.

Despite this discrepancy the voter model dynamics on small-world networks
is relatively insensitive to the quenched or annealed nature of the topology.
An interesting question for future work is whether this insensitivity
extends also to other ordering processes on other complex networks.

\appendix
\section*{Appendix}

In this Appendix we give explicit formulas for the functions
$f_{2 \nu} (p/L,L-2 \nu-1)$ and $f_{2 \nu+1}(p/L,L-2\nu-2)$.
Let us consider
\be
f_R (\alpha,N) \equiv \sum_{n=0}^{N}\left(\begin{array}{c} N \\
n \end{array}\right) \frac{\alpha^n (1-\alpha)^{N-n}}{R+n}.
\label{effe}
\ee
For $\alpha \neq 0$
\be
f_R (\alpha,N)=\frac{1}{\alpha^R}\int_0^{\alpha} ds s^{R-1} (s+1-\alpha)^N.
\label{inte}
\ee
The derivation of Eq.~(\ref{inte}) is easy
\bea
f_R (\alpha,N) &= & {1 \over \alpha^R}
\sum_{n=0}^{N}\left(\begin{array}{c} N \\ n \end{array}\right) 
\frac{\alpha^{n+R} (1-\alpha)^{N-n}}{R+n} \\ \nonumber
& & {1 \over \alpha^R}
\sum_{n=0}^{N}\left(\begin{array}{c} N \\ n \end{array}\right) 
(1-\alpha)^{N-n} \int_0^{\alpha} ds s^{R+n-1} \\ \nonumber
& & {1 \over \alpha^R} \int_0^{\alpha} ds s^{R-1}
\sum_{n=0}^{N}\left(\begin{array}{c} N \\ n \end{array}\right) 
(1-\alpha)^{N-n} s^n \\ \nonumber
& & {1 \over \alpha^R} \int_0^{\alpha} ds s^{R-1} (s+1-\alpha)^N. 
\eea
Using Eq.~(\ref{effe}) it is easy to verify that
\be
R f_R(\alpha,N-R) + \alpha (N-R) f_{R+1}(\alpha,N-R-1) = 1.
\label{eqapp}
\ee
The vanishing of the coefficient on the right hand side of
Eqs.~(\ref{M1}) and~(\ref{M2}) is obtained by setting $R=2\nu$,
$\alpha=p/L$ and $N=L-1$.

Using Eq.~(\ref{inte}) the explicit formulas for $f_2$ and $f_3$ are
readily found
\be
f_2(p/L,L-3) = \left({L \over p}\right)^2
\left[ {1 \over L-1} - {(1-p/L) \over L-2} + {(1-p/L)^{L-1} \over (L-2)(L-1)}
\right]
\ee
\be
f_3(p/L,L-4) = \left({L \over p}\right)^3
\left[ {1 \over L-1} - {2(1-p/L) \over L-2} + {(1-p/L)^2 \over L-3} -
{2(1-p/L)^{L-1} \over (L-3)(L-2)(L-1)}
\right].
\ee
Finally the asymptotic forms of the functions for $L \to \infty$ are
\be
F_2(p) = \lim_{L \to \infty} f_2(p/L,L-3) = {1 \over p} - 
{(1-e^{-p}) \over p^2}
\ee
and
\be
F_3(p) = \lim_{L \to \infty} f_3(p/L,L-4) = {1 \over p} - {2 \over p^2} +
{2(1-e^{-p}) \over p^3},
\ee
that go to $1/2$ and $1/3$, respectively, in the limit $p \to 0$.

\end{document}